\documentclass[aps,pra,twocolumn,superscriptaddress,preprintnumbers,
amsmath,amssymb,floatfix]{revtex4}

\usepackage{amsmath}
\usepackage{amsfonts}
\usepackage{amssymb}
\usepackage{graphicx}
\usepackage{color}
\usepackage{bm}
\usepackage{hyperref}
\usepackage[caption=false, font=small]{subfig}

\begin{document}

%
%

\newcommand{\Tr}{\mathrm{Tr}}
\newcommand{\ASlash}{\slashed{A}}
\newcommand{\pSlash}{\slashed{p}}
\newcommand{\epsSlash}{\slashed{\epsilon}}
\newcommand{\kSlash}{\slashed{k}}
\newcommand{\psiA}{\psi_{\mathcal{A}}}
\newcommand{\psiAmin}{\psiA^\text{min}}
\newcommand{\psiAmax}{\psiA^\text{max}}
\newcommand{\e}{\text{e}}
\renewcommand{\d}{\text{d}}
\renewcommand{\t}{\text{t}}
\renewcommand{\i}{\text{i}}
\renewcommand{\v}{\text{v}}
\renewcommand{\Re}{\text{Re}}
\renewcommand{\Im}{\text{Im}}


\newcommand{\be}{\begin{equation}}
\newcommand{\ee}{\end{equation}}
\newcommand{\beqa}{\begin{align}}
\newcommand{\eeqa}{\end{align}}
\newcommand{\beqan}{\begin{align*}}
\newcommand{\eeqan}{\end{align*}}
\newcommand{\ean}{\nonumber\\}
\newcommand{\vect}{\bm}
\newcommand{\PV}{\mathcal{C}\!\!\!\!\!\!\int}
\newcommand{\PVtext}{c\!\!\!\!\int}
\newcommand{\leftexp}[2]{{\vphantom{#2}}^{#1}{#2}}

\def\ket#1{$| #1 \rangle$}
\def\ketm#1{| #1 \rangle}
\def\bra#1{$\langle #1 |$}
\def\bram#1{\langle #1 |}
\def\spr#1#2{$\langle #1 | \right. #2 \rangle$}
\def\sprm#1#2{\langle #1 | \right. #2 \rangle}
\def\me#1#2#3{$\langle #1 | #2 | #3 \rangle$}
\def\mem#1#2#3{\langle #1 | #2 | #3 \rangle}
\def\redme#1#2#3{$\langle #1 \|
                  #2 \| #3 \rangle$}
\def\redmem#1#2#3{\langle #1 \|
                  #2 \| #3 \rangle}
\def\threej#1#2#3#4#5#6{$\left( \begin{matrix} #1 & #2 & #3  \\
                                               #4 & #5 & #6  \end{matrix} \right)$}
\def\threejm#1#2#3#4#5#6{\left( \begin{matrix} #1 & #2 & #3  \\
                                               #4 & #5 & #6  \end{matrix} \right)}
\def\sixj#1#2#3#4#5#6{$\left\{ \begin{matrix} #1 & #2 & #3  \\
                                              #4 & #5 & #6  \end{matrix} \right\}$}
\def\sixjm#1#2#3#4#5#6{\left\{ \begin{matrix} #1 & #2 & #3  \\
                                              #4 & #5 & #6  \end{matrix} \right\}}

\def\ninejm#1#2#3#4#5#6#7#8#9{  \left\{ \begin{matrix} #1 & #2 & #3  \\
                                                       #4 & #5 & #6  \\
                                                       #7 & #8 & #9  \end{matrix}  \right\}   }

%
%
%
%

%
%

\title{An x-ray quantum eraser setup for time-energy complementarity}

%
%

\author{Jonas \surname{Gunst}}
\email{Jonas.Gunst@mpi-hd.mpg.de}
\affiliation{Max-Planck-Institut f\"ur Kernphysik, Saupfercheckweg 1, D-69117 Heidelberg, Germany}

\author{Adriana \surname{P\'alffy}}
\email{Palffy@mpi-hd.mpg.de}
\affiliation{Max-Planck-Institut f\"ur Kernphysik, Saupfercheckweg 1, D-69117 Heidelberg, Germany}


\date{\today}

%
%
%
%
%
%
%
\begin{abstract}

A new quantum eraser setup exploiting for the first time the time-energy complementarity relation in the x-ray regime is investigated theoretically.
Starting point is the interference process between x-ray quanta driving two nuclear hyperfine transitions in a nuclear forward scattering setup.
We show that \textit{which-way} information can be obtained by marking the scattering paths with orthogonal polarization states, thus leading to the
disappearance of the interference pattern. In turn, erasure of the \textit{which-way} information leads to the reappearance of the interference fringes.
We put forward two schemes using resonant scattering off nuclear targets and design \textit{which-way} marking procedures to realize the quantum eraser setup for x-ray quanta.

\end{abstract}
%
%




\maketitle


\section{Introduction \label{intro}}

In quantum mechanics, complementarity refers to the principle stipulated by Niels Bohr back in 1927 \cite{Bohr1928} that classical measurements cannot reveal all information about a quantum system within a single experimental setup. Due to the 
inseparability between the quantum system and the classical detector, some observables become mutually exclusive, i.e., they are complementary. The precise knowledge of one of them implies that all possible outcomes of measuring the other observable are equally probable. Although originally examples of complementarity referred mostly to the position (particle-like) and momentum (wave-like) attributes of a quantum system, one can also connect for instance energy and time, or different spin projections. Complementarity is closely related to Heisenberg's uncertainty relations \cite{Heisenberg1927}  that describe the limitations on the possible accuracy of classical measurements of a quantum system. Over the years, the dilemmas of complementarity gedankenexperiments such as Einstein's recoiling-slit two-slit experiment \cite{BohrEinstein} or Feynman's electron-light scattering scheme \cite{Feynman} were solved invoking the position-momentum uncertainty relation \cite{Heisenberg1927}, prompting the question whether complementarity is always enforced by the uncertainty relations. A vivid theoretical debate \cite{Scully1991.N,Storey1994.N,replyScully,replyStorey,commentWiseman} as well as several experiments \cite{Eichmann1993,Duerr1998,Mir2007} showed that complementarity is more fundamental than the uncertainty relation, i.e., one can find cases in which the effect of Heisenberg's uncertainty relations is not sufficient to justify the occurring complementarity.

Quantum interference, the key concept in complementarity (gedanken)experiments, occurs whenever a quantum system can choose between several paths from a common initial state $|i\rangle$ to the same final state $|f\rangle$. The most famous interference experiment is Young's double-slit setup. As long as the path chosen by the particle through the double-slit remains unrevealed, one can observe interference fringes in a position measurement somewhere behind the slits. According to the complementarity principle, a detector capable of determining the path taken by the particle through the double-slit (and supply {\it which-way} information) will destroy the interference pattern. The physical mechanism via which the interference is lost typically 
 relies on random classical momentum kicks which are introduced by the detector revealing {\it which-way} information in the system \cite{BohrEinstein}. This is however not always the case. In more recent quantum optics experiments, 
the path chosen by the particle, this time an atom or ion, can be marked by  employing further degrees of freedom related to their internal, quantum structure. In this case, even if the acquisition of {\it which-way} information involves a momentum kick, this is a coherent quantum process rather than a classical random one. While the interference disappears due to this marking, one can later choose to ``erase'' the {\it which-way} information and restore the interference behavior, in a process first proposed in Ref.~\cite{eraser1982} and illustratively called {\it quantum eraser}. If the instant of erasure is chosen such that it occurs only after detection of the particle (allowing determination of its earlier behavior as wave or particle) one speaks of a delayed-choice quantum eraser \cite{delayed2000,delayed2014}.

The basic elements of a quantum eraser, as enumerated in Ref.~\cite{Zeilinger1995} are individual interfering quantum systems, a method of introducing which-path information, and a method of subsequently erasing this information in order to restore the interference. Various realizations of the quantum eraser in more or less traditional systems have been reported so far. First experiments in this direction employed entangled optical photon pairs for the study of interference \cite{Zeilinger1995} and momentum-position complementarity \cite{Ou1990,Zajonc1991,Kwiat1992}, while later on also more exotic systems have been employed or envisaged, for instance, mesoscopic systems \cite{Hans1998,mesoScience2014}, kaons \cite{kaons2004}, nuclear spins \cite{nmr2001}, continuous-variable quantum erasing using field quadrature amplitude and phases \cite{Leuchs2004} or ultrafast quantum emitters in microcavities \cite{Savasta2011}. We note that most of these setups address the position-momentum complementarity, with  few exceptions \cite{Zeilinger1995,kaons2004,Leuchs2004}. 

The time-energy complementarity relation has as counterpart the time-energy uncertainty relation, $\Delta E\Delta t\ge \hbar/2$. We recall that the latter, although generally accepted as valid, does not rely on the same mathematical basis as other uncertainty relations, i.e., it is not a direct 
mathematical consequence of the replacement of classical numbers by operators. Time is not an operator in quantum mechanics, it is just a parameter.
 Although from the theoretical and philosophical side the debates on the concept of time in quantum mechanics are ongoing \cite{Bohm1961,Kijowski1974,Muga2000,Vona2013}, there are only few experiments with quantum particles which do not involve time-related measurements. A number of double-slit setups recording interference between paths differing in time rather than spatially have been investigated both theoretically and experimentally. These include a double-slit experiment in the time-energy domain where the slits are related to different time windows of attosecond duration \cite{attoslit2005,sfaslit2006,attoslit2006},  a streaking temporal double-slit in an orthogonal two-color laser field \cite{Doerner2015}, vacuum-mediated interference in atomic resonant fluorescence \cite{Kiffner2006} or oscillations of M\"ossbauer neutrinos \cite{Akhmedov2009}.

In this work, we investigate theoretically a different quantum eraser setup exploiting the complementarity between time and energy in a so-far unexplored parameter regime. The novelty arises from the use of hard x-ray quanta instead of optical or infrared photons. Apart from x-ray detection efficiencies close to 100\%, shorter wavelengths open the possibility to achieve much better spatial resolutions down to the nanometer scale \cite{Mimura2010.NP}. However, x-rays are no longer resonant to valence electron transitions in atoms or ions, but rather to inner shell electron transitions in (highly) charged ions \cite{Young2010.N,Kanter2011.PRL,Rohringer2012}, or transitions in atomic nuclei \cite{Buervenich2006,rohlsberger2012electromagnetically,Kilian2015.PRA,Xiangjin2016.PRL-slowlight,Jonas2016.SciRep}. The latter bring the advantage to provide clean quantum mechanical systems well-isolated from their environment.

\begin{figure}%
\centering
\subfloat[NFS setup]{\label{fig:NFS-QE-1}\includegraphics[width=0.6\linewidth]{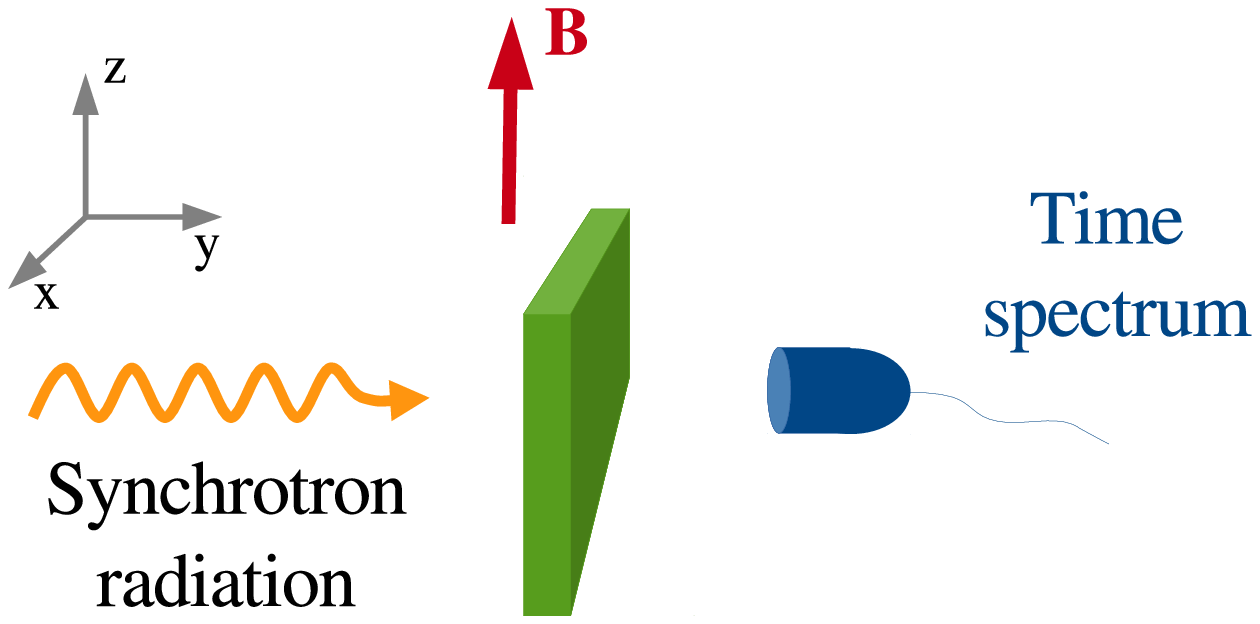}}%
\subfloat[Quantum beats]{\label{fig:NFS-QE-2}\includegraphics[width=0.4\linewidth]{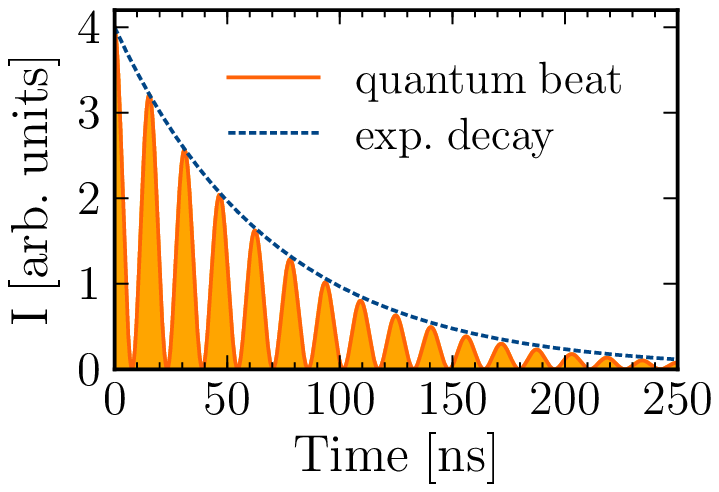}}\\%
\subfloat[Individual ${}^{57}$Fe nucleus]{\label{fig:NFS-QE-3}\includegraphics[width=0.66\linewidth]{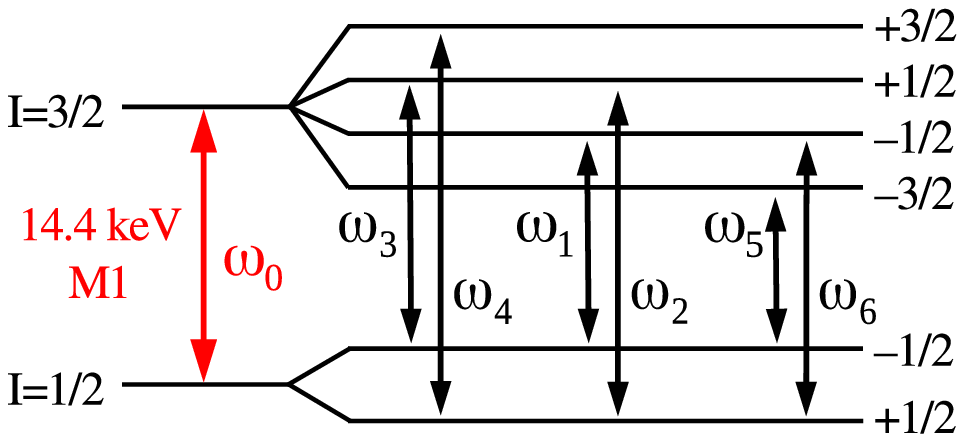}}%
\subfloat[Collective picture: $\Delta M=0$]{\label{fig:NFS-QE-4}\includegraphics[width=0.33\linewidth]{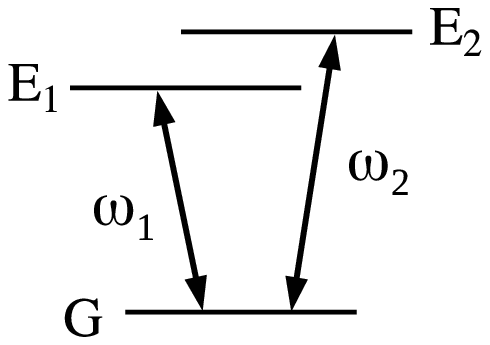}}%
\caption{(a) Typical NFS setup with magnetic field $\bm{B}$. (b) NFS time spectrum for a thin target with quantum beat pattern (orange curve). (c) Nuclear level scheme of an individual ${}^{57}$Fe nucleus under hyperfine splitting. (d) Collective picture with a common ground state $G$ and excited levels $E_1$ and $E_2$ driven via the $\Delta M=0$ transitions.}
\label{fig:NFS-QE}
\end{figure}

We envisage interference occurring in driving the 14.4 keV M\"ossbauer transition in $^{57}\mathrm{Fe}$ by x-rays from a synchrotron radiation (SR) source. Under the presence of a hyperfine magnetic field, the magnetic splitting of the ground and first excited states will lead to the indistinguishable excitation of two transitions as shown in Fig.~\ref{fig:NFS-QE} and to the detection of an interference pattern. Subsequent manipulations of the x-ray photon polarization in a marking procedure leads to the disappearance of the interference. By appropriate quantum erasure of the which-way information, the interference pattern can be restored in a very similar manner as explained in the original proposal \cite{Scully1991.N}. Due to their long coherence times, the use of nuclear transitions leads to very small relative values for the energy transfer occurring in the marking procedure compared to the actual photon energy, $E_{\rm{tr}}/E\sim 10^{-13}$. As it will be shown later on, an important role is  also played by the ratio between the nuclear transition width and the magnitude of the hyperfine splitting, $\Gamma_0/\Delta E$.

In order to obtain quantum interference  we exploit the collective effects rising in nuclear forward scattering (NFS) of SR on M\"ossbauer solid-state samples. Due to the recoilless excitation and decay of the nuclear excited state, as long as the initial and the final states are the same for all nuclei in the sample, it is impossible to pin-point which nucleus or nuclei were involved in the excitation. In this respect, we have here the generalized, $N$-scatterer version of Young's interference experiment with (optical) light 
scattered from two atoms \cite{Eichmann1993}. The collective excitation leads to  directional scattering in the forward direction allowing a spatial separation of the coherent (collective) and incoherent (without interference) decays \cite{NFS-bible}. The collective coherent decay and consequently the interference effects occur in the forward direction. This high directionality is strongly related to the ability of M\"{o}ssbauer nuclei to absorb and re-emit photons without recoil, which is hardly realizable in atomic systems. We present two versions of a quantum eraser setup where the originally interfering paths are first marked via orthogonal polarization states, leading to a wash-out of the interference fringes. 
By applying a projection to a different polarization basis which acts as a quantum eraser,  the which-way information is lost and the interference fringes reappear. We discuss the energy shifts 
implied by the polarization marking and the relation between complementarity and the uncertainty relation in our case.

The paper is structured as follows. Sec.~\ref{sec:NFS} introduces the concept of NFS and the theoretical formalism used to describe the interference spectra in the collective decay. Sec.~\ref{sec:quantum-eraser} continues with the description of the polarization marking techniques and describes two versions of  quantum erasure for the discussed setups. Numerical results are presented here. The paper concludes with a summary and discussion of the results.

\captionsetup[subfigure]{position=top}

\section{Nuclear forward scattering}
\label{sec:NFS}

\subsection{Interference mechanisms}
\label{sec:NFS-interference}

A typical NFS setup is shown in Fig.~\ref{fig:NFS-QE-1}. The x-rays, typically from a SR source monochromatized to the nuclear transition energy, propagate along the $y$-direction and impinge on the nuclear sample with an incident angle of $90^{\circ}$. The radiation is linearly polarized with $x$-($z$-)polarized light denoted as $\sigma$-($\pi$-)polarization by convention \cite{Siddons1999.HI}.  The time spectrum of the resonantly scattered radiation is detected in the forward direction [see Fig.~\ref{fig:NFS-QE-2}]. The nuclear response occurs on a much longer time scale than the x-ray pulse duration and the non-resonant, electronic response, allowing for time gating of the signal \cite{Roehlsberger2004}. Due to the typically narrow nuclear resonances and the low brilliance of x-ray sources, at most one nucleus can be excited in the sample per pulse. 

The most used nuclear transition with the NFS technique is the one connecting the stable ground state of $^{57}$Fe (nuclear spin $I_g=1/2$) with the the first excited state (nuclear spin $I_e=3/2$, mean lifetime $\tau$=141~ns) at 14.413 keV, as shown in Fig.~\ref{fig:NFS-QE-3}.  The recoilless nature of this transition in solid-state nuclear targets leads to the formation of a delocalized, collective excitation which decays coherently into the forward direction leading to a relative speed-up and enhancement of the NFS yield. The formation of the exciton can be interpreted as an interference effect for the scattering of the resonant photon or photons off a collection of $N$ nuclei. As long as the identity of the interacting nucleus is not revealed, for instance via spin flip, recoil or an internal conversion electron, the collective excitation occurs.

In the presence of a nuclear hyperfine magnetic field, the ground and excited nuclear states undergo Zeeman splitting according to their spin values as illustrated in Fig.~\ref{fig:NFS-QE-3}. Due to the Fourier limit of the temporally narrow incident x-ray pulses, several polarization-selected hyperfine transitions can be simultaneously driven  leading to well-known quantum beats in the NFS intensity spectrum \cite{NFS-bible}. For instance, initially $\sigma$-($\pi$-)polarized x-rays couple to all $\Delta M=0$ ($\Delta M=\pm1$) transitions provided the magnetic field $\bm{B}$ points along the $z$-direction. Since only those photons are coherently scattered into the forward direction for which the nucleus returns  to its original ground state Zeeman level, the $\sigma$-($\pi$-)polarization is conserved in the course of NFS with constant hyperfine field $\bm{B}$. In such a setup, the quantum mechanical state of a single nucleus can in general be expressed as the following superposition
\begin{align}
\ketm{\Psi} =& \sum_{i=1}^2 \Big( c_{e_i} \ketm{e_i,0} + c_{g_i} \ketm{g_i,0} \Big) + c_1 \ketm{g_1, 1_{\omega_1}} \nonumber \\
&\qquad + c_2 \ketm{g_2, 1_{\omega_2}}\ ,
\end{align}
where \ket{g_i,0} and \ket{e_i,0} ($i=1,2$) represent the states in ground and excited magnetic sublevels, respectively, with the radiation field in vacuum, \ket{g_i, 1_{\omega_i}} ($i=1,2$) denote the states after the reemission process in the course of NFS involving one photon with either frequency $\omega_1$ or $\omega_2$, and $c_{g_i}$, $c_{e_i}$ and $c_{i}$ ($i=1,2$) are the corresponding probability coefficients. According to the full quantum mechanical description, the two $\Delta M=0$ transitions do not interfere for a single nucleus due to the different initial and final states involved. The beat pattern between $\omega_1$ and $\omega_2$ evaluates in this case to
\begin{align}
\mem{\Psi}{E_1^{(-)} E_2^{(+)}}{\Psi} 
& \propto \mem{g_1, 1_{\omega_1}}{a_1^{\dagger} a_2 \e^{\i(\omega_1-\omega_2)t}}{g_2, 1_{\omega_2}} \nonumber \\
& \propto \mem{1_{\omega_1}}{a_1^{\dagger} a_2}{1_{\omega_2}}  \e^{\i(\omega_1-\omega_2)t} \langle g_1 | g_2 \rangle \nonumber \\
& \propto \e^{\i(\omega_1-\omega_2)t} \underbrace{\langle g_1 | g_2 \rangle}_{=0} \ ,
\end{align}
where the field operators $E_i^{(\pm)}$ are proportional to $a_i \e^{-\i \omega_i t}$ and $a_i^{\dagger}\e^{\i \omega_i t}$, respectively, with $a_i$ and $a_i^{\dagger}$ being the photon creation and annihilation operators belonging to frequency $\omega_i$ ($i=1,2$). The beat pattern disappears for a single nucleus because the Zeeman levels $g_1$ and $g_2$ are orthogonal to each other, $\langle g_1 | g_2 \rangle=0$.

In an ensemble of many nuclei collective effects come into play. In this case, the two $\Delta M=0$ transitions are connected by the collective ground state which contains nuclei on both hyperfine levels,
\begin{equation}
 | G \rangle =\underbrace{| g_{1}^{(1)}\rangle \ldots |g_{1}^{(N_1)} \rangle}_{N_1} \underbrace{| g_{2}^{(N_1+1)}\rangle \ldots |g_{2}^{(N)} \rangle}_{N_2}\ ,
\label{ground}
\end{equation}
where $| g_{1}\rangle$ and $| g_{2}\rangle$ denote the two ground magnetic sublevels and $N_i$ is the number of nuclei on the ground state $| g_{i}\rangle \, (i\in{1,2})$ with $N_1+N_2=N$. At room temperature, typically $N_1\approx N_2$.
The excited state on the other hand can be described as a timed Dicke state  \cite{Scully2009.superradiance}
\begin{equation}
 | E_{\mu} \rangle =\frac{1}{\sqrt{N_{\mu}}}\sum_{n=1}^{N_{\mu}}\e^{\i\bm{k} \cdot \bm{r}_n}|g_{1}^{(1)}\rangle \ldots |e_{\mu}^{(n)} \rangle \ldots | g_{2}^{(N)}\rangle\ ,
\label{excited}
\end{equation}
in which the $n^{\rm{th}}$ nucleus has been excited via the transition $\mu$, with the notation $\mu=1$ for the transition $M_g=-1/2\rightarrow M_e=-1/2$ and $\mu=2$ for $M_g=+1/2\rightarrow M_e=+1/2$, depending on the initial ground state spin projection $M_g$. The position of the excited nucleus is given by $\bm{r}_n$ and $\bm{k}$ represents the wave vector of the resonant incident x-ray field. The scattering channels with $\Delta M=0$ are equivalent in this system to the  transitions $|G\rangle \rightarrow |E_1\rangle$ and $|G\rangle \rightarrow |E_2\rangle$, as illustrated in Fig.~\ref{fig:NFS-QE-4}. 

We show in the following that interference effects can occur in this setup (see Ref.~\cite{ScullyZubairy}). Having at most one excitation inside the nuclear ensemble, the general state of the system can be written as
\begin{align}
\ketm{\Psi_{\rm{coll}}} =& \sum_{i=1}^2 C_{E_i} \ketm{E_i,0} + C_{G} \ketm{G,0} + C_1 \ketm{G, 1_{\omega_1}} \nonumber \\
&\qquad + C_2 \ketm{G, 1_{\omega_2}}
\end{align}
with $C_{E_i}$, $C_G$ and $C_i$ representing the probability amplitudes for being in \ket{E_i,0}, \ket{G,0} and \ket{G_i,1_{\omega_i}} ($i=1,2$), respectively. Following the same steps as in the case of a single nucleus, we obtain
\begin{align}
\mem{\Psi_{\rm{coll}}}{E_1^{(-)} E_2^{(+)}}{\Psi_{\rm{coll}}} 
& \propto \mem{G,1_{\omega_1}}{a_1^{\dagger} a_2 \e^{\i(\omega_1-\omega_2)t}}{G,1_{\omega_2}} \nonumber \\
& \propto \e^{\i(\omega_1-\omega_2)t} \underbrace{\langle G | G \rangle}_{=1}\ .
\end{align}
Here, we observe a beat pattern since the two $\Delta M=0$ transitions are connected via the common ground state \ket{G}.
This is equivalent with the coherent addition of classical fields, that is introduced in the following.

\subsection{Theoretical approach}
\label{sec:NFS-theory}

The approach we are using follows Refs.~\cite{Shvydko1999.HI, Shvydko1999.PRB, Shvydko2000.HI} where Maxwell's equations
are directly solved in time and space. This kind of treatment allows us to easily incorporate time-dependent interactions like sudden switchings of the magnetic field. Actually, Ref.~\cite{Shvydko1999.PRB} provides a general way of solving the problem of NFS in terms of a semiclassical description, applicable to arbitrary time-dependent interactions. Here, we rather present a reduced version of the model, adapted to our purposes.

\subsubsection{Wave equation}

We describe the coherent nuclear scattering process with a semiclassical approach where the electromagnetic field is considered to behave classically whereas the nuclear system is treated quantum mechanically. 
In order to facilitate the treatment of time-dependent hyperfine interactions later, it is natural to solve the scattering problem directly in space and time based on Refs.~\cite{Shvydko1999.HI, Shvydko1999.PRB, Shvydko2000.HI}.
Accordingly, the coherent propagation of the electromagnetic field through the resonant medium is thereby described by Maxwell's equation, 
\begin{equation}
\frac{\partial}{\partial y} \bm{E}(y,t) = -\frac{2 \pi}{c} \bm{J}(y,t),
\label{eq:NFS_Maxwell-slowly}
\end{equation}
where the slowly-varying amplitude approximation has been applied. Here, the amplitudes $\bm{E}(y,t)$ and $\bm{J}(y,t)$ denote the electromagnetic field inside the nuclear target and the macroscopic current density induced by the incident radiation, respectively. These amplitudes only depend on the spatial coordinate $y$ because absorption as well as refraction occur along the propagation direction. The boundary condition for $\bm{E}(y,t)$ is set by the incident radiation field,
\begin{equation}
\bm{E}(0,t) = \bm{E}_{\rm{in}}(t)\ .
\label{eq:NFS_Maxwell-boundary}
\end{equation}

Once we have solved Eq.~(\ref{eq:NFS_Maxwell-slowly}), the forward scattered intensity spectrum behind the target is given by
\begin{equation}
I(t) = \left| \bm{E}(L,t) \right|^2\ ,
\label{eq:NFS_intensity}
\end{equation}
where $L$ denotes the thickness of the nuclear sample.

\subsubsection{General solution}

In order to solve Eq.~(\ref{eq:NFS_Maxwell-slowly}) it is helpful to introduce dimensionless space and time variables $\xi$ and $\tau$, respectively, via
\begin{align}
\xi &= \frac{1}{4} \sigma_{\rm{R}} n_0 y\ , \nonumber \\
\label{eq:NFS_xi-tau}
\tau &= \Gamma_0 t\ ,
\end{align}
where $n_0$ is the number of resonant nuclei per unit volume, $\sigma_{\rm{R}}$ the total nuclear cross section of resonant absorption and $\Gamma_0$ the full natural transition width. For the special case of $y=L$, the spatial variable $\xi$ is also known as effective thickness of the resonant scattering.

Using these definitions and writing the macroscopic nuclear current density as a sum over all individual nuclear current densities \cite{Shvydko1999.HI}, it is possible to express Eq.~\eqref{eq:NFS_Maxwell-slowly} in the following form
\begin{equation}
\frac{\partial}{\partial \xi} \bm{E}(\xi,\tau) = -\sum_{\ell} \bm{J}_{\ell}(\bm{k},\tau) \int_{-\infty}^{\tau} \d\tau'\, \bm{J}_{\ell}^{*}(\bm{k},\tau') \cdot \bm{E}(\xi,\tau')\ .
\label{eq:NFS_wave-eq-dimensionless}
\end{equation}
The occurring summation runs over all contributing hyperfine transitions $\ell$, e.g., the ones depicted in Fig.~\ref{fig:NFS-QE-3} for ${}^{57}$Fe. Each nuclear excitation and de-excitation via a certain transition $\ell$ is described by the matrix elements $\bm{J}_{\ell}^*$ and $\bm{J}_{\ell}$, respectively.
The resulting wave equation \eqref{eq:NFS_wave-eq-dimensionless} is similar to a Schr\"{o}dinger equation where time is replaced by the dimensionless space variable $\xi$. Analogously to perturbation theory it is possible to represent the solution as a power series in $\xi$,
\begin{equation}
\bm{E}(\xi,\tau) = \sum_{p=0}^{\infty} \bm{E}^{(p)}(\xi,\tau) = \sum_{p=0}^{\infty} \frac{(-\xi)^p}{p!} \bm{E}^{(p)}(\tau)\ .
\label{eq:NFS_scattering-ampl}
\end{equation}
The term $p=0$ represents the boundary condition given in Eq.~\eqref{eq:NFS_Maxwell-boundary} and reads
\begin{equation}
\bm{E}^{(0)}(\xi,\tau) = \bm{E}_{\rm{in}}(\tau)\ .
\label{eq:NFS_scattering-ampl-boundary}
\end{equation}
All higher order terms are obtained from power matching in $\xi$ by inserting Eq.~(\ref{eq:NFS_scattering-ampl}) into the wave equation \eqref{eq:NFS_wave-eq-dimensionless}, leading to
\begin{equation}
\bm{E}^{(p)}(\tau) = \sum_{\ell} \bm{J}_{\ell}(\bm{k},\tau) \int_{-\infty}^{\tau} \d\tau'\, \bm{J}_{\ell}^{*}(\bm{k},\tau') \cdot \bm{E}^{(p-1)}(\tau')\ .
\label{eq:NFS_scattering-ampl-solution}
\end{equation}

Since the first order is in most cases the dominating scattering contribution (in particular for thin samples or in general for small interaction times), it is useful to write down the explicit form of $\bm{E}^{(1)}$,
\begin{equation}
\bm{E}^{(1)}(\xi,\tau) = -\xi \sum_{\ell} \bm{J}_{\ell}(\bm{k},\tau) \left( \bm{J}_{\ell}^{*}(\bm{k},0) \cdot \bm{e}_{\rm{p}}\right)\ ,
\label{eq:NFS_scattering-ampl-1st-delta}
\end{equation}
where the incident radiation was considered to be a $\delta$-like pulse in time with arbitrary polarization $\bm{e}_{\rm{p}}$.

\subsubsection{Static hyperfine field}

In the case of static hyperfine interactions, it is possible to express the time-dependent nuclear transition currents $\bm{J}_{\ell}^{*}$ and $\bm{J}_{\ell}$ in terms of the time-independent matrix elements of the current density operator $\bm{j}_{\ell}^{*}$ and $\bm{j}_{\ell}$, respectively. Taking the explicit time evolution of the nuclear spin states into account, we obtain
\begin{align}
\bm{J}_{\ell}^*(\bm{k},\tau) &= \e^{ \i \Omega_{\ell} \tau - \tau/2 } \bm{j}_{\ell}^*(\bm{k})\ , \nonumber \\
\label{eq:NFS_nucl-current-static}
\bm{J}_{\ell}(\bm{k},\tau) &= \e^{ -\i \Omega_{\ell} \tau - \tau/2 } \bm{j}_{\ell}(\bm{k})\ ,
\end{align}
with
\begin{equation}
\Omega_{\ell} = \left( M _{\rm{g}} \epsilon _{\rm{g}} - M _{\rm{e}} \epsilon _{\rm{e}} \right) / \Gamma_0
\label{eq:NFS_omegal}
\end{equation}
describing the frequency correction due to magnetic hyperfine splitting in units of $\Gamma_0$. Here, we assumed that the magnetic field points along the $z$-direction, $\bm{B}=B_0 \bm{e}_z$, and introduced the factor $\epsilon_{\lambda}=\mu_{\lambda} B_0/I_{\lambda}$ ($\lambda \in \{\rm{g},\rm{e}\}$) standing for the energy splitting caused by the hyperfine interactions, where the nuclear states are characterized by their total spin quantum number $I_{\lambda}$ with projections $M_{\lambda}$ and magnetic moments $\mu_{\lambda}$.
For instance, in the case of ${}^{57}$Fe the two $\Delta M=0$ resonances are energetically separated by $\Delta E= |\epsilon _{\rm{g}} - \epsilon _{\rm{e}}|$.
The amplitudes $\bm{j}_{\ell}^*$ and $\bm{j}_{\ell}$ appearing in Eqs.~(\ref{eq:NFS_nucl-current-static}) denote the time-independent transition elements of the current density operators $\bm{j}^{\dagger}$ and $\bm{j}$, respectively. It is important to note that the directionality of $\bm{j}_{\ell}^*$ and $\bm{j}_{\ell}$ is strongly related to the polarization of the incident light and the direction of the magnetic field \cite{Shvydko1999.PRB,Adriana2010.JoMO}. 

Plugging the explicit time dependence of the nuclear currents (\ref{eq:NFS_nucl-current-static}) into Eq.~(\ref{eq:NFS_scattering-ampl-1st-delta}), the first order solution for NFS in the presence of a static hyperfine field can be written as
\begin{equation}
\bm{E}^{(1)}(\xi,\tau) = -\xi \sum_{\ell} \bm{\mathcal{A}}^{(1)}_{\ell,\mathrm{I}}(\bm{k}) \, \e^{-\i \Omega_{\ell} \tau - \tau/2}\ .
\label{eq:field-I}
\end{equation}
Accordingly, the single scattering events can be expressed as a coherent summation over all contributing nuclear transitions $\ell$ where the summands factorize into a time-dependent phase factor $\e^{-\i \Omega_{\ell} \tau - \tau/2}$ and a time-independent amplitude $\bm{\mathcal{A}}^{(1)}_{\ell,\mathrm{I}}$ which is determined by
\begin{equation}
\bm{\mathcal{A}}^{(1)}_{\ell,\mathrm{I}}(\bm{k}) = \bm{j}_{\ell}(\bm{k}) \left( \bm{j}_{\ell}^{*}(\bm{k}) \cdot \bm{e} _{\rm{p}} \right)\ .
\label{eq:amplitude-I}
\end{equation}

\subsubsection{Fast switchings of the magnetic field}

In this paragraph the hyperfine interactions investigated so far are generalized to the case where it is allowed to switch the magnetic field off and on, as already discussed in Ref.~\cite{Wente2012-storage} in the context of coherent storage of x-ray photons. We consider therefore an initially applied $\bm{B}$-field to be switched off at time $\tau=\tau_0$ and on again at $\tau=\tau_1$. Thereby, the magnetic field after $\tau_1$ is considered to point in the same direction than the initial one.

Analogously to the case of static hyperfine interactions, it is intended to express the nuclear currents $\bm{J}_{\ell}$ in terms of the time-independent transition amplitudes $\bm{j}_{\ell}$. The only difference in the calculation is that the time evolution of nuclear states splits up into three time sectors: $\mathrm{I})$ $\tau<\tau_0$, $\mathrm{II})$ $\tau_0<\tau<\tau_1$ and $\mathrm{III})$ $\tau>\tau_1$. In sector $\mathrm{I}$, the nuclear currents are again given by Eqs.~\eqref{eq:NFS_nucl-current-static} and the first order solution reduces to Eq.~\eqref{eq:field-I}.

In sector $\mathrm{II}$, the magnetic field strength is zero and the Zeeman levels become degenerated. Depending on the moment $\tau_0$ when the magnetic field is turned off, we obtain
\begin{align}
\label{eq:NFS_nucl-current-switch-II}
\bm{J}_{\ell,\mathrm{II}}(\bm{k},\tau) &= \e^{-\i \Omega_{\ell} \tau_0 - \tau/2} \bm{j}_{\ell}(\bm{k})\ .
\end{align}
Using the explicit form of the nuclear currents the first order scattering solution for $\tau_0<\tau<\tau_1$ evaluates to
\begin{equation}
\bm{E}^{(1)}(\xi,\tau) = -\xi \sum_{\ell} \bm{\mathcal{A}}^{(1)}_{\ell,\mathrm{II}}(\bm{k}) \, \e^{- \tau/2}\ ,
\label{eq:field-II}
\end{equation}
with $\bm{\mathcal{A}}^{(1)}_{\ell,\mathrm{II}} = \bm{\mathcal{A}}^{(1)}_{\ell,\mathrm{I}}\, \e^{-\i \Omega_{\ell} \tau_0}$.

In sector $\mathrm{III}$, the analogous procedure is applied resulting in the following expression for the nuclear currents
\begin{align}
\label{eq:NFS_nucl-current-switch-III}
\bm{J}_{\ell,\mathrm{III}}(\bm{k},\tau) &= \e^{-\i \Omega_{\ell} \tau_0 - \tau/2} \sum_{\ell'} \e^{-\i \Omega_{\ell'} (\tau - \tau_1)} \bm{j}_{\ell'}(\bm{k})\ .
\end{align}
Inserting this expression into Eq.~\eqref{eq:NFS_scattering-ampl-1st-delta}, leads to the first order scattering solution for $\tau>\tau_1$,
\begin{equation}
\bm{E}^{(1)}(\xi,\tau) = -\xi \sum_{\ell} \bm{\mathcal{A}}^{(1)}_{\ell,\mathrm{III}}(\bm{k}) \, \e^{-\i \Omega_{\ell} \tau - \tau/2}\ ,
\label{eq:field-III}
\end{equation}
where the time-independent amplitudes $\bm{\mathcal{A}}^{(1)}_{\ell,\mathrm{III}}$ are given by 
\begin{equation}
\bm{\mathcal{A}}^{(1)}_{\ell,\mathrm{III}}(\bm{k}) = \bm{j}_{\ell}(\bm{k}) \, \e^{\i \Omega_{\ell} \tau_1} \sum_{\ell'} \, \e^{-\i \Omega_{\ell'} \tau_0} \left( \bm{j}_{\ell'}^{*}(\bm{k}) \cdot \bm{e} _{\rm{p}} \right)\ .
\label{eq:amplitude-III}
\end{equation}

\section{Which-way information and quantum eraser}
\label{sec:quantum-eraser}

The occurrence of the quantum beat pattern in the course of NFS can be explained in analogy to a double-slit setup. Instead of the interference of two spatial path ways like in a conventional double-slit experiment, the quantum beat pattern in the course of NFS is caused by the interplay between the frequency paths contributing to the scattering process, e.g., $\omega_{1,\dots,6}$ in the case of ${}^{57}$Fe [see Fig.~\ref{fig:NFS-QE-3}]. In the following we will restrict ourselves to the interference of the two $\Delta M=0$ transitions, $\omega_1$ and $\omega_2$.
We put forward two versions of a quantum eraser setup. The general idea of both is to first mark the originally interfering paths via orthogonal polarization states, leading to a wash-out of the interference fringes. By applying a projection to a different polarization basis which acts as a quantum eraser, the which-way information is lost and the interference fringes reappear.

\subsection{Scheme 1: Two targets in a collinear setup}

\begin{figure}
\centering
\includegraphics[width=1.0\linewidth]{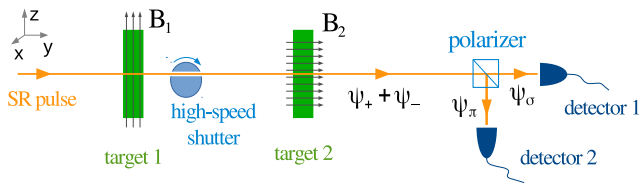}
\caption{Quantum-eraser scheme 1. The $\sigma$-polarized synchrotron pulse is resonantly scattered at target 1 ($\bm{B}_1 \parallel \bm{e}_z$) via the two $\Delta M=0$ transitions. The high-speed shutter cuts the SR pulse off such that the incoming radiation field at target 2 is solely the nuclear response of target 1. Following the nuclear scattering off target 2 ($\bm{B}_2 \parallel \bm{e}_y$) a polarizer splits the $\sigma$- and $\pi$-polarization components.}
\label{fig:QE-two-targets}
\end{figure}

The setup of scheme 1 is presented in Fig.~\ref{fig:QE-two-targets}, consisting of two ${}^{57}$Fe targets with magnetic fields $\bm{B}_1$ and $\bm{B}_2$, respectively, a high-speed shutter \cite{Toellner2011.shutter} behind the first target and a polarizer directly in front of the detectors. Similar two-target setups, however without magnetic fields have been experimentally demonstrated in Refs.~\cite{helistoe1982,helistoe1991,Smirnov1996NE, Smirnov2005}. In the following, we will describe the propagation of the SR pulse through this setup step by step. The essential building blocks of a quantum eraser will be identified and the required conditions evaluated.

The incident SR pulse is considered to be $\sigma$-polarized initially and the magnetic field at target 1 is chosen to point along the $z$-direction such that only the two $\Delta M=0$ transitions are driven. For the considered $\bm{B}$-field geometry the $\sigma$-polarization is conserved in the course of the nuclear resonance scattering. Within the applied hyperfine field, target 1 adopts essentially the role of the double slit in the original quantum eraser proposal \cite{Scully1991.N}, imprinting the quantum beat pattern on the scattered field.

\begin{figure}%
\centering
\subfloat[Target 1]{\label{fig:QE-scheme2-1}\includegraphics[width=0.5\linewidth]{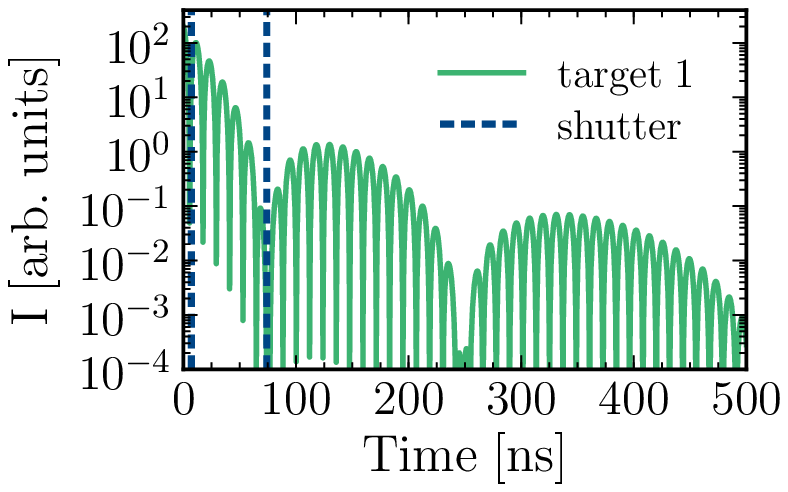}}%
\subfloat[Shutter + Target 2]{\label{fig:QE-scheme2-2}\includegraphics[width=0.5\linewidth]{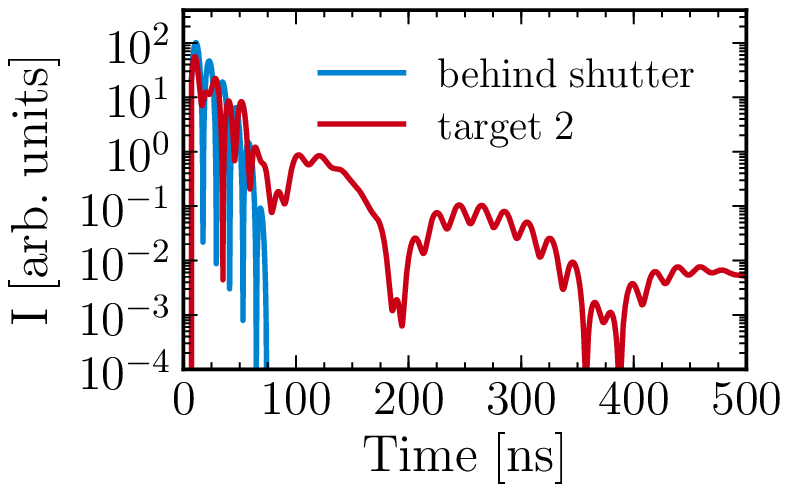}}%
\caption{(a) NFS intensity spectrum $|\bm{E}(\xi,\tau)|^2$ directly behind target 1.
The scattering via the two $\Delta M=0$ transitions results in the characteristic quantum beat pattern. Due to the shutter only photons emitted in the time window from 7~ns until 74~ns [region between blue dashed lines in (a), blue curve in (b)] reach target 2. (b) The magnetic field $\bm{B}_2$ is chosen such that the interference between the $\Delta M=0$ frequency components is destroyed in the intensity spectrum $|\bm{E}(\xi,\tau)|^2$ (red curve) directly behind target 2.
Effective thicknesses of $\xi_1=\xi_2=7$ and a maximal scattering order of $p_{\rm{max}}=19$ have been considered in the calculations.}
\label{fig:QE-scheme2-marker}
\end{figure}

Fig.~\ref{fig:QE-scheme2-1} shows the time spectrum of the forward scattered intensity directly behind target 1. Due to the two scattering paths in frequency space the quantum beat interference pattern appears in the time spectrum. Apart of the interference between the $\Delta M=0$ channels, the dynamical beat modulation caused by multiple scattering events inside the sample occurs in the intensity spectrum. In the calculation presented in Fig.~\ref{fig:QE-scheme2-1}, an effective target thickness for sample 1 of $\xi_1=7$ has been considered.

\captionsetup[subfigure]{position=top}

The role of the shutter behind target 1 is to select a certain time window of the scattered field. This time window can be controlled by the shutter properties like the diameter, the width of the slit or the position of the latter, and by the applied rotation frequency \cite{Toellner2011.shutter}. In Fig.~\ref{fig:QE-scheme2-2} for instance, we assumed that only photons emitted from $t_0=7$~ns until $t_1=74$~ns are able to pass the shutter and to reach the second target. The choice of $t_0>0$ also cuts off the broadband SR pulse which behaves essentially $\delta$-like in time in comparison to the nuclear response. In frequency space, the high-speed shutter creates two spectrally narrow x-ray pulses at frequencies $\omega_1=\omega_0-\Omega_2 \Gamma_0$ and $\omega_2=\omega_0+\Omega_2 \Gamma_0$, respectively, by inverting the two $\Delta M=0$ absorption dips. 
Note that due to the broadband nature of SR light and the narrow nuclear transition widths, typically at most a single resonant x-ray photon is involved in the scattering process, either with frequency $\omega_1$ or $\omega_2$.

The magnetic field at the second iron target points along the $y$-axis which is the direction of propagation. This is known as the Faraday geometry and in this case only $\Delta M=\pm1$ transitions are allowed according to angular momentum conservation. Photons emitted via the $\Delta M=+1$ and $\Delta M=-1$ are right and left circularly polarized, respectively. The idea to delete the interference pattern imprinted at target 1 is to tune the strengths of the magnetic fields $\bm{B}_1$ and $\bm{B}_2$ such that the spectrally narrow peaks impinging on target 2 are marked with orthogonal circular polarizations. One choice is to set the conditions $\omega_1^{(1)} = \omega_3^{(2)}$ and $\omega_2^{(1)} = \omega_6^{(2)}$, resulting in
\begin{equation}
\frac{B_1}{B_2} = \frac{\mu_{\rm{g}}/I_{\rm{g}}+\mu_{\rm{e}}/I_{\rm{e}}}{\mu_{\rm{g}}/I_{\rm{g}}-\mu_{\rm{e}}/I_{\rm{e}}}\ ,
\end{equation}
where $\mu_{\rm{g}}$ and $\mu_{\rm{e}}$ denote the magnetic moments of the ground and the excited states, respectively. In terms of the nuclear magneton $\mu_{\rm{n}}$, the magnetic moments of ${}^{57}$Fe are given by $\mu_{\rm{g}}=0.09044\, \mu_{\rm{n}}$ and $\mu_{\rm{e}}=-0.1549\, \mu_{\rm{n}}$ \cite{ENSDF}. In the same manner, one can impose $\omega_1^{(1)} = \omega_5^{(2)}$ and $\omega_2^{(1)} = \omega_4^{(2)}$ which can be realized by
\begin{equation}
\frac{B_1}{B_2} = \frac{\mu_{\rm{g}}/I_{\rm{g}}-3\mu_{\rm{e}}/I_{\rm{e}}}{\mu_{\rm{g}}/I_{\rm{g}}-\mu_{\rm{e}}/I_{\rm{e}}}\ .
\end{equation}
However, for both cases a small chance remains to drive off-resonant transitions, e.g. $\omega_4^{(2)}$ and $\omega_5^{(2)}$ in case 1, or $\omega_3^{(2)}$ and $\omega_6^{(2)}$ in case 2, respectively. Perfect cancellation of the quantum beat pattern can be achieved only if these off-resonant interactions are totally negligible.

The intensity spectrum behind target 2 shown in Fig.~\ref{fig:QE-scheme2-2} is calculated with Zeeman splittings of $\epsilon_{\rm{g}}^{(1)} \approx 48\, \Gamma_0$ and $\epsilon_{\rm{e}}^{(1)} \approx -27\, \Gamma_0$ at target 1, and $\epsilon_{\rm{g}}^{(2)} \approx 28\, \Gamma_0$ and $\epsilon_{\rm{e}}^{(2)} \approx -16\, \Gamma_0$ at target 2, corresponding to the case where $\omega_1^{(1)} = \omega_5^{(2)}$ and $\omega_2^{(1)} = \omega_4^{(2)}$. Moreover, an effective thickness of $\xi_2=7$ has been considered. As can be seen from Fig.~\ref{fig:QE-scheme2-2}, for photons interacting with both targets the which-way information is successfully obtained, destroying the interference pattern. The condition of double interaction is certainly fulfilled at times larger than $t_1$ such that the quantum beat completely disappears in this region. Only the dynamical beat signature remains in the intensity spectrum except for small oscillations coming from interactions with off-resonant transitions.

\begin{figure}%
\centering
\subfloat{\label{fig:QE-scheme2-3a}\includegraphics[width=0.519\linewidth]{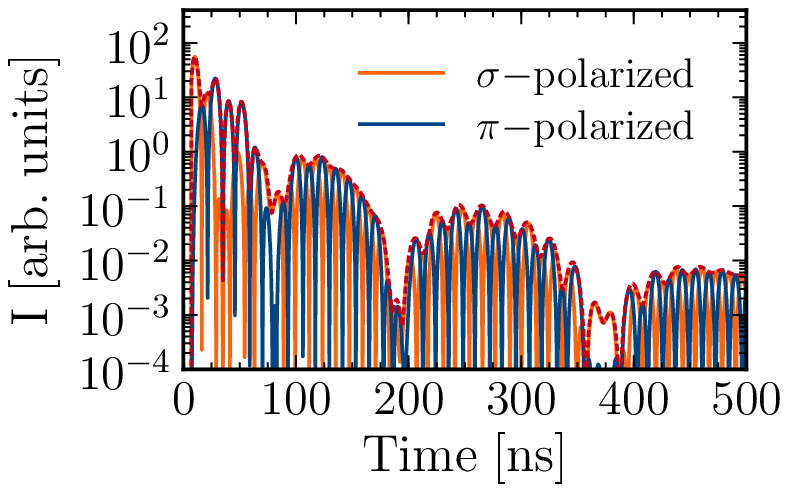}}%
\subfloat{\label{fig:QE-scheme2-3b}\includegraphics[width=0.481\linewidth]{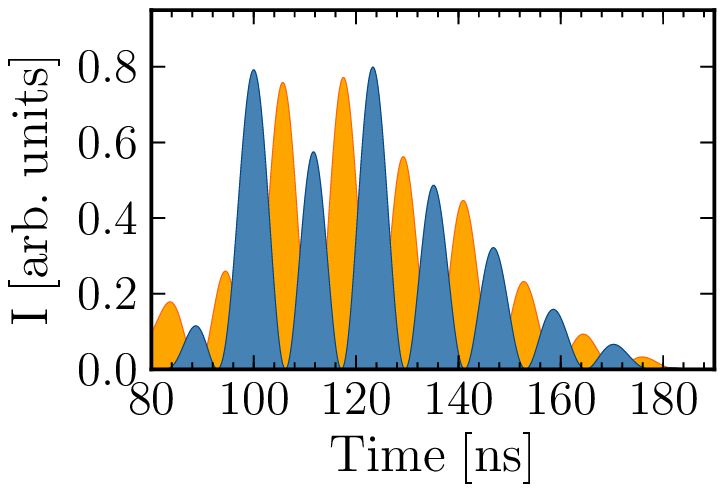}}
\caption{The $\sigma$- (orange curve) and $\pi$-components (blue curve) of the intensity spectrum $|\bm{E}(\xi,\tau)|^2$ are shown together with the unpolarized signal (red dotted line)  from Fig.~\ref{fig:QE-scheme2-marker}. While the left graph covers the region from 0 to 500~ns, the right graph zooms into the time window between 80 and 190~ns. In the calculations scattering orders up to $p_{\rm{max}}=19$ have been taken into account.}
\label{fig:QE-scheme2-eraser}
\end{figure}

One may wonder at this point whether the disappearance of the interference pattern is simply due to a change in energy of the individual scattering paths introduced by the marking procedure. We recall that this was the topic of a vivid argument 
on uncertainty over complementarity for quantum eraser experiments in the momentum-space domains in Refs.~\cite{Scully1991.N, Storey1994.N, replyScully, replyStorey, commentWiseman}. Also in our case, we have to admit that the marking procedure might apply an energy ``kick'' to the scattered photon: each resonant absorption and reemission of an x-ray photon introduces an energy uncertainty on the order of the natural transition width $\Gamma_0$ leading to a correspondingly uncertain emission time. In turn, the period of the quantum beat, i.e., the magnitude of the interference fringe,  is determined by the hyperfine splitting $\Delta E$. The resolution of the quantum beats may therefore be diminishing the larger the ratio $\Gamma_0/\Delta E$ leading to a wash-out of the interference pattern \cite{vanBurck1999.PRA}. We would like however to point out that although an energy ``kick'' is applied in the marking procedure, this cannot be the complete reason why the interference pattern is  disappearing. Let us consider for the moment a collinear setup with two targets both under the action of the same magnetic field. Also in this case, the absorption and reemission of the x-ray photon in the second target introduces an energy uncertainty on the order of the natural transition width $\Gamma_0$, but since no marking via orthogonal polarizations has occurred, the interference pattern persists. This has been discussed in Refs.~\cite{Siddons1999.HI, Wente2012-storage}. 
Thus, there is more to obtaining which-way information than mere random energy ``kicks'' to justify the disappearance of interference. This is why it is also possible to restore at a latter point the interference pattern by erasing the which-way information.

After the successful marking of the scattering paths with orthogonal polarizations the question arises, how to erase again the which-way information obtained by the resonant scattering off the second target? Since the two scattering paths are marked with circular polarizations, a polarizer projecting on the linear polarization basis $\bm{e}_{\sigma}$ and $\bm{e}_{\pi}$ destroys the orthogonality and restores the interference pattern as shown in Fig.~\ref{fig:QE-scheme2-eraser}. The left graph of Fig.~\ref{fig:QE-scheme2-eraser} presents the intensity spectrum at detector 1 (orange curve) and detector 2 (blue curve) over the whole time range employed in Fig.~\ref{fig:QE-scheme2-marker}, whereas the right graph of Fig.~\ref{fig:QE-scheme2-eraser} zooms into the time region between 80 ns and 190 ns. In the latter, the relative shift between the $\sigma$- and the $\pi$-components of a quarter of a beating period can be clearly seen. The $\sigma$- and $\pi$-polarized parts take the role of the fringes and anti-fringes introduced in Ref.~\cite{Scully1991.N}.

The small oscillations in Fig.~\ref{fig:QE-scheme2-2} show that the marking via orthogonal polarizations can be slightly distorted due to the off-resonant interactions. In order to minimize the latter, the incoming spectrally narrow peaks as well as the absorption lines need to be well separated in frequency space. This corresponds to shrinking the ratio $\Gamma_0/\Delta E$ as much as possible, e.g., by using a very large magnetic splitting $\Delta E$ which leads to a large detuning for the off-resonant transitions. A large magnetic splitting  $\Delta E$ in turn involves high magnetic fields.
In the calculations presented in Figs.~\ref{fig:QE-scheme2-marker} and \ref{fig:QE-scheme2-eraser}, for instance, magnetic field strengths of $B_1\approx39$~T and $B_2\approx23$~T have been considered in order to obtain the intended hyperfine splittings.
Since such high magnetic field strengths are difficult to realize in laboratory,
it is worth to explore if a quantum eraser scenario is also feasible in a setup where a strong internal hyperfine field can be employed, e.g. by using antiferromagnetic FeBO$_3$ crystals.

\subsection{Scheme 2: Interferometer-like setup}

Instead of using two targets in a sequence, it is also possible to realize a quantum eraser scheme by employing an x-ray interferometer setup involving two spatially separated targets, one in each interferometer arm. The general idea is to directly apply a time delay in one of the arms such that a relative phase is imprinted. Considering for instance a geometry where only the two $\Delta M=0$ transitions, $\omega_1$ and $\omega_2$, are driven, it turns out that a time delay in one of the arms produces a phase shift for each frequency component exactly opposite in sign. How this phase shift can be employed to destroy and recover the quantum beat pattern of NFS is explained in the following.

\begin{figure}
\centering
\includegraphics[width=0.9\linewidth]{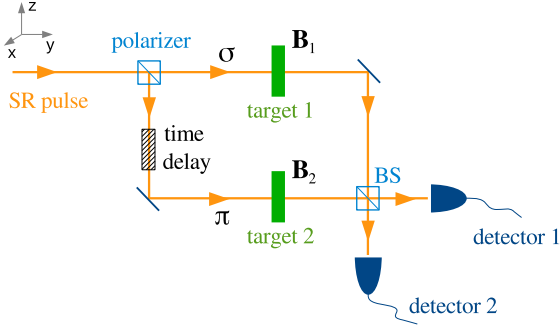}
\caption{Quantum-eraser scheme 2. The incoming x-ray photon is split into its $\sigma$- and $\pi$-polarized components by a polarizer \cite{Toellner1995.polarizer,Marx2013.polarizer}. The split beams are later on mixed with the help of two mirrors \cite{Shvydko2011.NP} and a beam splitter (BS). In each arm a nuclear ${}^{57}$Fe target is inserted. The two targets experience the magnetic fields $\bm{B}_1$ and $\bm{B}_2$, respectively. A relative time delay between arm 1 and 2 is introduced.}
\label{fig:QE-interferometer}
\end{figure}

The x-ray interferometer we are considering is schematically presented in Fig.~\ref{fig:QE-interferometer}. In comparison to a conventional Mach-Zehnder-interferometer, a peculiarity here is the polarizer at the beginning which splits the $\sigma$-polarized component from the $\pi$-polarized part.
After the polarizer, the split beam interacts with two nuclear samples, one in each interferometer arm, in the presence of the magnetic fields $\bm{B}_1$ and $\bm{B}_2$, respectively. The $\pi$-polarized pulse furthermore experiences a time delay $\Delta\tau$ in comparison to the $\sigma$-polarized counterpart, before reaching target 2. Finally, both spatially separated arms come together at a beam splitter (BS) \cite{Osaka2013.OE} where they are recombined.

The interferometer system composed of a polarizer, two mirrors, two spatially separated nuclear targets and finally a beam splitter (BS) as shown in Fig.~\ref{fig:QE-interferometer} can be theoretically described in the following way \cite{Agarwal}
\begin{align}
\begin{pmatrix}
\bm{E}_{\rm{out-1}}^{(1)} \\
\bm{E}_{\rm{out-2}}^{(1)}
\end{pmatrix}
=&
\frac{1}{2}
\underbrace{
\begin{pmatrix}
1 & \i \\
\i & 1
\end{pmatrix}
}_{\textrm{``BS''}}
\underbrace{
\begin{pmatrix}
-1 & 0 \\
0 & -1
\end{pmatrix}
}_{\textrm{``mirrors''}}
\underbrace{
\begin{pmatrix}
\psi_1 & 0 \\
0 & \psi_2
\end{pmatrix}
}_{\textrm{``targets''}}
\nonumber \\
&\qquad \times
\underbrace{
\begin{pmatrix}
\ketm{\sigma}\bram{\sigma} & 0 \\
0 & \ketm{\pi}\bram{\pi}
\end{pmatrix}
\begin{pmatrix}
1 & \i \\
\i & 1
\end{pmatrix}
}_{\textrm{``polarizer''}}
\begin{pmatrix}
\bm{E}_{\rm{in}} \\
0
\end{pmatrix}\ .
\label{eq:QE-interferometer}
\end{align}
Here, $\bm{E}_{\rm{in}}$ represents the incoming synchrotron pulse and $\bm{E}_{\rm{out-1}}^{(1)}$ and $\bm{E}_{\rm{out-2}}^{(1)}$ the output at detector 1 and 2, respectively, in the single scattering approximation.
Considering the synchrotron pulse initially in an arbitrary linear polarization state, as given by the superposition 
\begin{equation}
\bm{E}_{\rm{in}}(\tau) = \left( \alpha \bm{e}_{\sigma} + \beta \bm{e}_{\pi} \right) \delta(\tau),
\end{equation}
with $\alpha,\beta \in \mathbb{R}$ and normalization $\alpha^2+\beta^2=1$, and performing the matrix multiplications occurring in Eq.~\eqref{eq:QE-interferometer}, we obtain 
\begin{align}
\begin{pmatrix}
\bm{E}_{\rm{out-1}}^{(1)} \\
\bm{E}_{\rm{out-2}}^{(1)}
\end{pmatrix}
=
\frac{1}{2}
\begin{pmatrix}
-\alpha{\bm{E}^{(1)}}^{\sigma}(\xi,\tau) + \beta{\bm{E}^{(1)}}^{\pi}(\xi,\tau) \\
-\i \alpha{\bm{E}^{(1)}}^{\sigma}(\xi,\tau) - \i \beta{\bm{E}^{(1)}}^{\pi}(\xi,\tau)
\end{pmatrix}\ .
\label{eq:E-out}
\end{align}
Here, ${\bm{E}^{(1)}}^{\sigma}$ and ${\bm{E}^{(1)}}^{\pi}$ denote the first order contributions originating from the initially $\sigma$- and $\pi$-polarized component, respectively. Important to remark is that we only consider thin targets in this section such that the single scattering approximation is always the dominating contribution of the forward scattered field. Moreover, the time delay $\Delta\tau$ was not yet been accounted for in Eqs.~\eqref{eq:QE-interferometer} and \eqref{eq:E-out}. It is later on incorporated into the nuclear response ${\bm{E}^{(1)}}^{\pi}$ coming from target 2.

So far, the description in Eq.~\eqref{eq:E-out} is rather general, neither the incident radiation pulse nor the $\bm{B}$-field geometry has been fixed. Since we consider two spatially separated nuclear targets, the magnetic fields at each individual sample can be chosen independently of each other.
In the following, we now first choose the magnetic fields $\bm{B}_1$ and $\bm{B}_2$ such that only the $\Delta M=0$ transitions are driven in both targets. Explicitly, this can be achieved by a magnetic field $\bm{B}_1$ at target 1 parallel to the $z$-axis and $\bm{B}_2$ at target 2 parallel to the $x$-axis. The magnitudes of both are furthermore assumed to coincide. In order to evaluate the terms ${\bm{E}^{(1)}}^{\sigma}$ and ${\bm{E}^{(1)}}^{\pi}$ corresponding to the scattered field from each individual target, we employ the description of NFS introduced in Sec.~\ref{sec:NFS-theory} where a semi-classical wave equation is used. Moreover, we label the transitions $M_g=-1/2\rightarrow M_e=-1/2$ and $M_g=+1/2\rightarrow M_e=+1/2$ in the following via $\ell=-2$ and $\ell=2$, respectively. According to Eq.~\eqref{eq:field-I}, the first order solutions for the considered geometry can be written as
\begin{align}
{\bm{E}^{(1)}}^{\sigma}(\xi,\tau) &= -\xi \Big\{ \bm{\mathcal{A}}^{{(1)}^{\mbox{\scriptsize$\sigma$}}}_{\ell=-2,\mathrm{I}}(\bm{k}) \e^{\i \Omega_{2} \tau} + \bm{\mathcal{A}}^{{(1)}^{\mbox{\scriptsize$\sigma$}}}_{\ell=2,\mathrm{I}}(\bm{k}) \e^{-\i \Omega_{2} \tau} \Big\} \nonumber \\
&\qquad \times \e^{-\tau/2} 
\label{eq:E-target1}
\end{align}
and
\begin{align}
{\bm{E}^{(1)}}^{\pi}(\xi,\tau) &= -\xi \Big\{ \bm{\mathcal{A}}^{{(1)}^{\mbox{\scriptsize$\pi$}}}_{\ell=-2,\mathrm{I}}(\bm{k}) \e^{\i \Omega_{2} (\tau-\Delta\tau)} + \bm{\mathcal{A}}^{{(1)}^{\mbox{\scriptsize$\pi$}}}_{\ell=2,\mathrm{I}}(\bm{k}) \nonumber \\
&\qquad \times \e^{-\i \Omega_{2} (\tau-\Delta\tau)} \Big\} \e^{-(\tau-\Delta\tau)/2}  \ ,
\label{eq:E-target2}
\end{align}
where the amplitudes $\bm{\mathcal{A}}^{{(1)}^{\mbox{\scriptsize$\sigma$}}}_{\ell,\mathrm{I}}$ and $\bm{\mathcal{A}}^{{(1)}^{\mbox{\scriptsize$\pi$}}}_{\ell,\mathrm{I}}$ are defined in Eq.~\eqref{eq:amplitude-I} with $\bm{e}_{\rm{p}}=\bm{e}_{\sigma}$ and $\bm{e}_{\rm{p}}=\bm{e}_{\pi}$, respectively. Using the expressions for the nuclear hyperfine currents $\bm{j}_{\ell}^*$ and $\bm{j}_{\ell}$ \cite{Shvydko1999.PRB,Adriana2010.JoMO}, the amplitudes can be explicitly evaluated. Furthermore, the time delay $\Delta\tau$ is incorporated into the scattered field ${\bm{E}^{(1)}}^{\pi}$ originating from target 2 [see Eq.~\eqref{eq:E-target2}].

Inserting the NFS solutions of the individual targets [Eqs.~\eqref{eq:E-target1} and \eqref{eq:E-target2}] into Eq.~\eqref{eq:E-out}, leads to
\begin{align}
\bm{E}_{\rm{out-1}}^{(1)}(\xi,\tau)
& = -\xi \frac{f_{\rm{LM}}}{4} \Big\{ \e^{\i \Omega_{2} \tau} (-\alpha\bm{e}_{\sigma} + \beta' \e^{\i \Phi} \bm{e}_{\pi})
+ \e^{-\i \Omega_{2} \tau} \nonumber \\
&\qquad \times (-\alpha\bm{e}_{\sigma} + \beta' \e^{-\i \Phi} \bm{e}_{\pi}) \Big\} \e^{-\tau/2} \ ,
\nonumber \\
\bm{E}_{\rm{out-2}}^{(1)}(\xi,\tau)
&= \i\xi \frac{f_{\rm{LM}}}{4} \Big\{ \e^{\i \Omega_{2} \tau} (\alpha\bm{e}_{\sigma} + \beta' \e^{\i \Phi} \bm{e}_{\pi})
+ \e^{-\i \Omega_{2} \tau} \nonumber \\
&\qquad \times (\alpha\bm{e}_{\sigma} + \beta' \e^{-\i \Phi} \bm{e}_{\pi}) \Big\} \e^{-\tau/2} \ , 
\label{eq:E-out-general}
\end{align}
where we have introduced the phase $\Phi=\Omega_2 \Delta\tau$ and the coefficient $\beta'=\beta \e^{\Delta\tau/2}$. Moreover, $f_{\rm{LM}}$ represents the so-called Lamb-M\"{o}ssbauer factor describing the probability for a recoilless scattering event. As seen from Eqs.~\eqref{eq:E-out-general}, the $\pi$ component experiences a relative phase shift of $\Phi$ in comparison to the $\sigma$-polarized part in the case of $\ell=-2$, while the phase shift is of opposite sign for $\ell=2$.

Let us first discuss the simplest scenario with zero time delay, $\Delta\tau=0$. In this case $\beta'$ reduces to $\beta$ and the phase $\Phi$ is identically zero such that the factor $\e^{\i\Phi}$ evaluates to one. The polarization response of each frequency component follows the superposition state of the incident radiation $\bm{E}_{\rm{in}}$. Taking the modulus squared of $\bm{E}_{\rm{out-1}}^{(1)}$ and $\bm{E}_{\rm{out-2}}^{(1)}$, respectively, the intensity spectra at each detector have the same form given by,
\begin{equation}
I_{\rm{out}}^{(1)}(\xi,\tau)
= \xi^2 \frac{f_{\rm{LM}}^2}{8} \Big\{ 1 + \cos( 2\Omega_2 \tau ) \Big\} \e^{-\tau} \ .
\label{eq:I-out-zero-delay}
\end{equation}
As can be seen from Eqs.~\eqref{eq:I-out-zero-delay}, in the case of $\Delta\tau=0$ the quantum beat pattern represented by the cosine term is preserved in the intensity output. 
Since the interferometer arms are characterized by orthogonal polarizations in the considered scenario, there is no interference term between arm 1 and arm 2 occurring in the intensity spectra. For this reason, the signals at detectors 1 and 2 coincide and it is sufficient to discuss only one of the outputs whenever talking about intensities in the following.

In order to switch off the quantum interference between the two hyperfine transitions $\ell=-2$ and $\ell=2$, a nonzero time delay $\Delta\tau$ can be employed. The idea is to choose $\Delta\tau$ such that the two scattering channels in frequency space, $\ell=-2$ and $\ell=2$, are marked by orthogonal polarization states, e.g., left and right circularly polarized. Therefore, the condition $\frac{\beta'}{\alpha} \e^{\i\Phi} = \pm\i$ needs to be fulfilled. Since $\alpha$ and $\beta$ are chosen to be real, the only possible solution is to set $\Phi=(2n +1)\pi/2$ ($n=0,1,2,\dots$) and $\beta'=\alpha$. Choosing for instance $n=0$, the phase requirement $\Phi=\pi/2$ can be achieved by a time delay of $\Delta\tau = \pi/(2\Omega_2)$ corresponding to a quarter of the beating period. 
In order to get completely rid of the quantum beat, the initial polarization has to be slightly rotated away from $45^{\circ}$ in order to compensate for the exponential decay term $\e^{-\Delta\tau/2}$ ($\beta'=\alpha$).
Taking the normalization $\alpha^2 + \beta^2 = 1$ and the condition $\beta'=\alpha$ into account, it is possible to fix the incident polarization in dependence of the time delay $\Delta\tau$,
\begin{align}
\alpha(\Delta\tau) &= \sqrt{\frac{1}{\e^{-\Delta\tau} +1}}\ , \nonumber \\
\beta(\Delta\tau) &= \sqrt{\frac{1}{\e^{\Delta\tau} +1}}\ .
\label{eq:alpha-beta}
\end{align}

Specifying Eqs.~\eqref{eq:E-out-general} to the conditions $\Phi=\pi/2$ and $\beta'=\alpha$, leads to the following field amplitudes at detector 1 and 2, respectively,
\begin{align}
\bm{E}_{\rm{out-1}}^{(1)}(\xi,\tau)
& = -\xi \alpha \frac{f_{\rm{LM}}}{4} \Big\{ \e^{\i \Omega_{2} \tau} (-\bm{e}_{\sigma} + \i \bm{e}_{\pi})
+ \e^{-\i \Omega_{2} \tau} \nonumber \\
&\qquad \times (-\bm{e}_{\sigma} - \i \bm{e}_{\pi}) \Big\} \e^{-\tau/2} \ ,
\nonumber \\
\bm{E}_{\rm{out-2}}^{(1)}(\xi,\tau)
& = \i\alpha \xi \frac{f_{\rm{LM}}}{4} \Big\{\e^{\i \Omega_{2} \tau} (\bm{e}_{\sigma} + \i \bm{e}_{\pi})
+ \e^{-\i \Omega_{2} \tau} \nonumber \\
&\qquad \times (\bm{e}_{\sigma} - \i \bm{e}_{\pi}) \Big\} \e^{-\tau/2} \ .
\label{eq:E-out-QE}
\end{align}
Since the phase shift $\Phi$ is of opposite sign for the two frequency slits $\ell=-2$ and $\ell=2$, each slit is marked by orthogonal polarization states, in Eqs.~\eqref{eq:E-out-QE} for instance, $\bm{e}_{-}$ for $\ell=-2$ and $\bm{e}_{+}$ for $\ell=2$ in the case of $\bm{E}_{\rm{out-1}}$, and vice versa in the case of $\bm{E}_{\rm{out-2}}$.  
As the orthogonal polarizations store the which-way information of the scattering process, the interference between the $\Delta M=0$ hyperfine transitions should vanish in the intensity spectra at the detectors. In order to prove this, we calculate $I_{\rm{out-1}}^{(1)}$ and $I_{\rm{out-2}}^{(1)}$ in the same manner as for zero time delay, resulting in
\begin{equation}
I_{\rm{out}}^{(1)}(\xi,\tau)
= \xi^2 \frac{f_{\rm{LM}}^2}{8} \e^{-\tau} \ .
\label{eq:I-out-QE}
\end{equation}
In comparison to Eqs.~\eqref{eq:I-out-zero-delay}, the quantum beats represented by the cosine function disappeared as expected.

\captionsetup[subfigure]{position=top}
\begin{figure*}%
\centering
\subfloat[Quantum beat]{\label{fig:QE-initial}\includegraphics[width=0.25\linewidth]{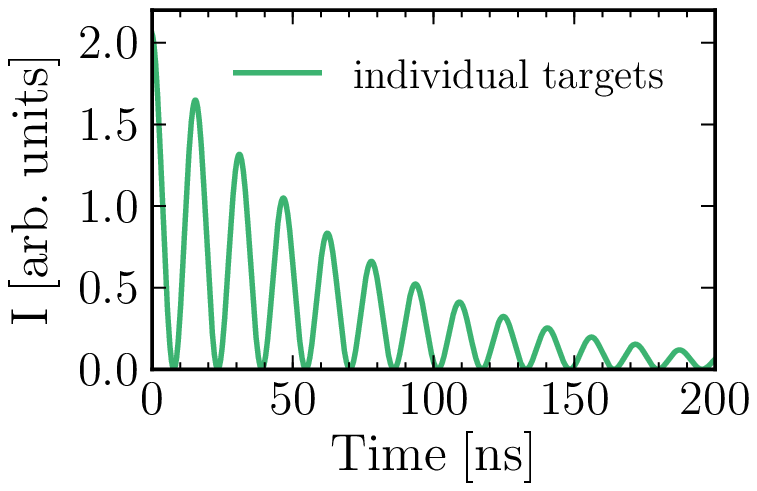}}%
\subfloat[Which-way, $\Phi=\frac{\pi}{2}$]{\label{fig:QE-inter}\includegraphics[width=0.25\linewidth]{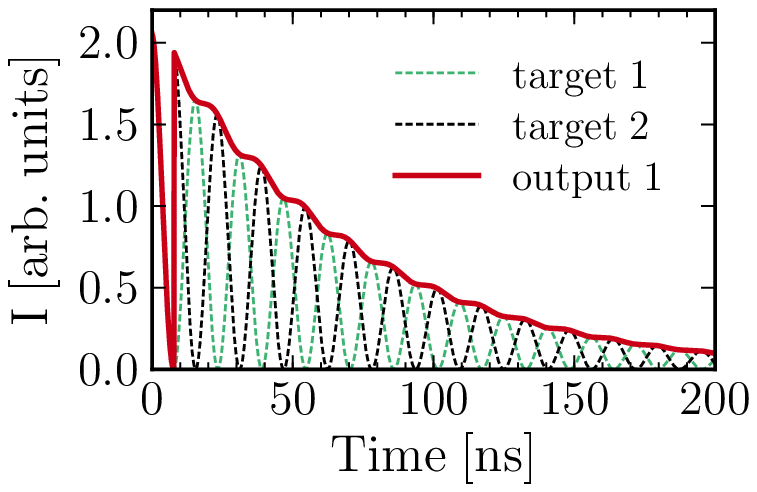}}%
\subfloat[Erasing]{\label{fig:QE-final}\includegraphics[width=0.25\linewidth]{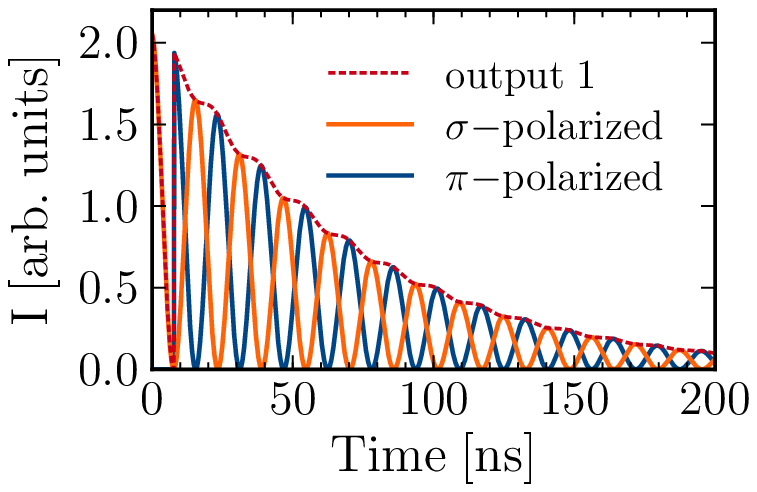}}%
\caption{Quantum eraser via a relative time delay. (a) The individual targets exhibit the quantum beat pattern due to interference between the two $\Delta M=0$ transitions. (b) Employing a time delay corresponding to $\Phi=\pi/2$ marks the scattering paths with orthogonal polarizations, destroying the interference. (c) The which-way information is erased and the interference is recovered by projecting onto the linear polarization basis, $\bm{e}_{\sigma}$ and $\bm{e}_{\pi}$.}
\label{fig:QE-time-delay}
\end{figure*}

In Fig.~\ref{fig:QE-time-delay}, we show numerical results of the intensity spectra at detector 1 for the cases $\Phi=0$ [Fig.~\ref{fig:QE-initial}] and $\Phi=\pi/2$ [Fig.~\ref{fig:QE-inter}]. Scattering orders up to $p_{\rm{max}}=14$ have been included in the calculations. A Zeeman splitting of $\Omega_2 \approx 28\, [\Gamma_0]$ (corresponding to the internal hyperfine field in FeBO$_3$) has been considered, resulting in a time delay $\Delta t=7.8$~ns for the special case of $\Phi=\pi/2$. Furthermore, the numerical results correspond to an effective target thickness $\xi=1$ and an incident polarization characterized by $\alpha=0.717$ and $\beta=0.697$ determined via the condition $\beta'=\alpha$ for the case of $\Phi=\pi/2$. Fig.~\ref{fig:QE-initial} shows that the quantum interference between the frequency slits $\ell=-2$ and $\ell=2$ is preserved for $\Delta\tau=0$ as already pointed out in Eq.~\eqref{eq:I-out-zero-delay}.

In the case of $\Phi=\pi/2$, the frequency paths $\ell=-2$ and $\ell=2$ should not interfere anymore, because the marking via the orthogonal polarizations $\bm{e}_{-}$ and $\bm{e}_{+}$, respectively, contains the which-way information (in frequency space) of the scattering process [see Eqs.~\eqref{eq:E-out-QE}]. This behavior is clearly recovered in Fig.~\ref{fig:QE-inter} where the intensity spectrum is determined by a simple exponential decay (red curve) instead of the quantum beat interference pattern. 

In order to restore the quantum interference by erasing the which-way information, a projection on the linear polarization basis (for instance with a polarizer) can be employed behind the beam splitter. 
Projecting on the $\sigma$- and $\pi$-polarization basis erases the which-way information stored in the orthogonal polarizations $\bm{e}_{-}$ and $\bm{e}_{+}$, reproducing the quantum beat pattern. The resulting intensity spectra are presented in Fig.~\ref{fig:QE-final}. Following the lines of the quantum eraser concept \cite{Scully1991.N}, the relatively shifted intensity spectra for the $\sigma$ and $\pi$ components can be interpreted as fringes and anti-fringes.

So far, we considered the time delay $\Delta\tau$ would be introduced by an external element, e.g., a time-delay line \cite{Roseker2011.time-delay, Roseker2013.time-delay, roling2012, osaka2014}. Instead of using an external element, fast switchings of the magnetic field at target 2 can be employed to coherently store the incident field for a while \cite{Wente2012-storage}. The hyperfine interaction for the case of a magnetic field instantaneously turned off and on again is described in detail in Ref.~\cite{Wente2012-storage}. 
We consider the case where the magnetic field $\bm{B}_2$ is switched off at time $\tau_0$ and on again at $\tau_1$. For both $\tau<\tau_0$ as well as $\tau>\tau_1$, $\bm{B}_2$ points along the $x$-axis.
Specifying Eqs.~\eqref{eq:field-I}, \eqref{eq:field-II} and \eqref{eq:field-III} to this case and following the procedure as it has been employed for a sudden turn off of the magnetic field in Ref.~\cite{Wente2012-storage}, we finally obtain the following scattering output (in first order) coming from target 2,
\begin{widetext}
\begin{equation}
{\bm{E}^{(1)}}^{\pi}(\xi,\tau) =
\begin{cases} 
\, -\xi \frac{f_{\rm{LM}}}{2} \Big\{ \e^{\i \Omega_{2} \tau} +\e^{-\i \Omega_{2} \tau} \Big\} \e^{-\tau/2} \bm{e}_{\pi} & \quad\textrm{for $\tau<\tau_0$} \\
\, -\xi \frac{f_{\rm{LM}}}{2} \Big\{ \e^{\i \Omega_{2} \tau_0} +\e^{-\i \Omega_{2} \tau_0} \Big\} \e^{-\tau/2} \bm{e}_{\pi} & \quad\textrm{for $\tau_0<\tau<\tau_1$} \\
\, -\xi \frac{f_{\rm{LM}}}{2} \Big\{ \e^{\i \Omega_{2} (\tau-\tau_1+\tau_0)} +\e^{-\i \Omega_{2} (\tau-\tau_1+\tau_0)} \Big\} \e^{-\tau/2} \bm{e}_{\pi} & \quad\textrm{for $\tau>\tau_1$}
\end{cases} \ .
\label{eq:E-target2-switch}
\end{equation}
\end{widetext}

\captionsetup[subfigure]{position=top}
\begin{figure*}%
\centering
\subfloat[Quantum beat, $\Phi=0$]{\label{fig:QE-storage-0}\includegraphics[width=0.25\linewidth]{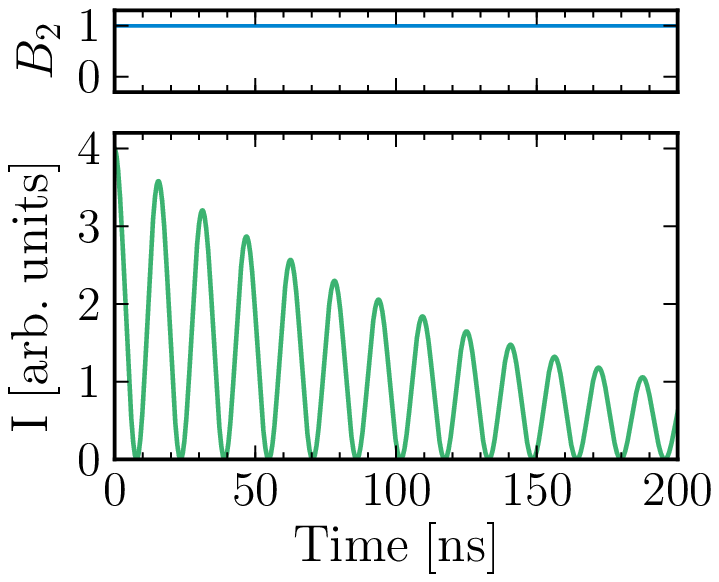}}%
\subfloat[Which-way, $\Phi=\frac{\pi}{2}$]{\label{fig:QE-storage-1}\includegraphics[width=0.25\linewidth]{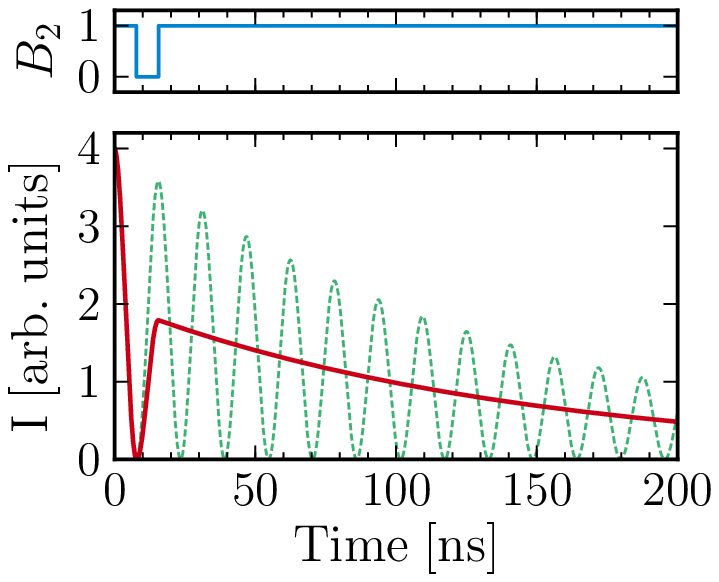}}%
\subfloat[Erasing, $\Phi=0$]{\label{fig:QE-storage-2}\includegraphics[width=0.25\linewidth]{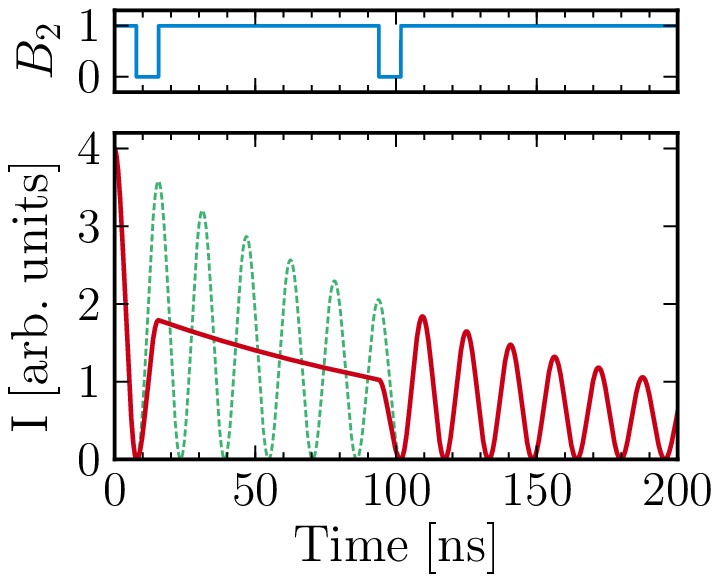}}%
\caption{Quantum eraser via an intrinsic storage scheme. The first-order intensity spectrum $|\bm{E}^{(1)}_{\rm{out-1}}|^2$ is shown along with the time sequence of magnetic field switchings. The initial quantum beats (a) are first annihilated via a relative phase shift $\Phi=\pi/2$ (b) and subsequently restored by erasing the which-way information via a second storage sequence, introducing an additional $\pi/2$ phase shift (c). }
\label{fig:QE-storage}
\end{figure*}

Turning the magnetic field instantaneously off at a time instant $\tau_0=(2n+1)\pi/(2\Omega_2)$ ($n=0,1,2,\dots$) which corresponds to a minimum of the quantum beat, results in a strongly suppressed emission at times $\tau_0<\tau<\tau_1$ where $\bm{B}_2=0$. After switching the magnetic field on again at $\tau_1$, the initial emission spectrum is recovered with an amplitude decreased by the exponential decay factor $\e^{-(\tau_1-\tau_0)/2}$. 
By using this storage scheme, a time delay $\Delta\tau = \tau_1-\tau_0$ (corresponding to the storage time) can be induced in comparison to the scattered field from target 1. Plugging Eq.~\eqref{eq:E-target1} and Eq.~\eqref{eq:E-target2-switch} into Eqs.~\eqref{eq:E-out}, leads for times $\tau>\tau_1$ qualitatively to the same behavior as an external time delay line as given in Eqs.~\eqref{eq:E-out-general} with the replacement $\beta' \mapsto \beta$. Choosing a setup with $\Delta\tau=\pi/(2\Omega_2)$ and an incident linear polarization of $45^{\circ}$ ($\alpha=\beta=1/\sqrt{2}$) marks the two frequency slits ($\ell=-2$ and $\ell=2$) by orthogonal polarization states, accomplishing the destruction of the interference pattern as shown in Fig.~\ref{fig:QE-storage-1}. In order to restore the interference ability between the two scattering path ways, it is possible to either make use of a polarizer to project on the $\sigma$- and $\pi$-polarization basis as done in Fig.~\ref{fig:QE-final} or to apply a second magnetic field switching introducing another relative time delay of $\pi/(2\Omega_2)$. The latter scenario is illustrated in Fig.~\ref{fig:QE-storage-2} where the cancellation of the interference pattern and its subsequent recovery can be followed in dependence on the sequence of magnetic field switchings.


\section{Summary and discussion}

In this article we propose two quantum eraser schemes potentially shifting time-energy complementarity tests towards so-far unexplored parameter regimes in the hard x-ray domain. This implementation  using for the first time nuclear systems instead of atoms would allow to test the universality of the quantum eraser concept  for a new energy regime and degree of complexity. The essential idea of both schemes is to cancel the quantum interference between two nuclear hyperfine transitions by marking the scattering paths with orthogonal polarizations. The knowledge of the \textit{which-way} information leads to the cancellation of the quantum beat pattern in the NFS time spectrum. By using a polarizer projecting on the linear polarization basis, the \textit{which-way} information can be erased and the beat pattern is restored.

In the first scheme we make use of a nuclear forward scattering setup with two collinear targets and a high-speed shutter in between. In general, the presence of magnetic fields would lead to a quantum beat pattern in the NFS intensity spectrum caused by the interference of simultaneously driven nuclear hyperfine transitions. Our results show that by choosing the right $\bm{B}$-field strengths and directions at target 1 and 2, it is possible to mark the scattering paths in frequency space by right- and left-circular polarizations, respectively, given that the scattered x-ray photon interacted with both targets. In this case the quantum beat disappears; however, it can be later on restored with the help of a polarizer. Similarly to the debate on uncertainty over complementarity for quantum eraser experiments in the momentum-space uncertainty relation \cite{Scully1991.N, Storey1994.N, replyScully, replyStorey, commentWiseman}, it appears that also in the case of energy and time,
the disappearance of the quantum beat pattern can be related but cannot be fully explained by small energy changes or ``kicks'' of the transition frequency in the marking procedure. A quantum eraser experiment in the x-ray regime could prove that these ``kicks'' are not random and the interference pattern can be restored by simply erasing the gathered which-way information. According to Ref.~\cite{commentWiseman}, this would show that complementarity is a more fundamental concept than the uncertainty principle. Finally, experimentally scheme 1 is challenging because of two reasons: (i) the marking procedure involves high magnetic field strengths, e.g., $B_1\approx39$~T and $B_2\approx23$~T as assumed in Fig.~\ref{fig:QE-scheme2-marker}, which additionally should be tunable over a certain parameter region; (ii) the magnetic field at target 2 is considered to be in the so-called Faraday geometry pointing in the propagation direction of the incident pulse.  

In the second scheme we consider an interferometer-like setup with one ${}^{57}$Fe target in each interferometer arm and a polarization-sensitive element at the entrance (see Fig.~\ref{fig:QE-interferometer}). A time delay $\Delta \tau$ in one of the interferometer arms leads to a relative phase shift between the resonantly driven $\Delta M =0$ transitions. In the case of an external delay element, our analysis shows that choosing $\Delta \tau$ and the incident polarization in dependence of the Zeeman splitting, the which-way information of the scattering process can be gained. We have furthermore shown that the external delay element can be replaced by a sequence of sudden magnetic field switchings leading to a similar phase shift between the contributing frequency components. By choosing appropriate switching times the marking as well as the erasing steps of a quantum eraser can be achieved. 
In contrast to an external time delay line, the intrinsic photon storage requires no adjustment of the incident polarization away from 45$^{\circ}$ [see Eqs.~\eqref{eq:alpha-beta}] to accomplish the erasing scheme. Moreover, time delays longer than a few ns seem to be easier feasible and additional degrees of control abilities like $\bm{B}$-field rotations are accessible, in comparison to external time delay lines. However, in order to apply switchings of the magnetic field at all, special iron samples without intrinsic Zeeman splitting like stainless steel are necessary. Turning magnetic fields of a few Tesla rapidly off and on may be realized according to Ref.~\cite{thesis-Wente} by two methods: (i) high-voltage snapper capacitors can be employed to regulate the pulse currents in magnetic coils; (ii) with the help of the lighthouse effect \cite{Roehlsberger2004} the nuclear sample can be quickly moved out of the region where the magnetic field is applied. In contrast, the scenario with an external time delay element does not require any switching schemes, opening the possibility to make use of the high intrinsic hyperfine field occurring in magnetized FeBO$_3$ targets.


\bibliographystyle{apsrev-no-url-issn}
\bibliography{eraser}{}


\end{document}